\documentclass[aps,prl,twocolumn,amsmath,showpacs,superscriptaddress]{revtex4}

\usepackage{graphicx}
\usepackage{amssymb}
\usepackage{pifont}

\def\mathbi#1{\textbf{\em #1}}

\begin{document}

\title{Ultrafast electro-nuclear dynamics of H$_{2}$ double ionization}

\author{S\'ebastien Saugout}
\affiliation{Laboratoire de Photophysique Mol\'{e}culaire du CNRS, Universit\'{e} Paris-Sud, B\^{a}timent 210, F-91405 Orsay, France.}
\affiliation{Service des Photons, Atomes et Mol\'{e}cules, Direction des Sciences de la Mati\`{e}re, CEA Saclay, B\^{a}timent 522, F-91191 Gif-sur-Yvette, France.}

\author{Christian Cornaggia}
\affiliation{Service des Photons, Atomes et Mol\'{e}cules, Direction des Sciences de la Mati\`{e}re, CEA Saclay, B\^{a}timent 522, F-91191 Gif-sur-Yvette, France.}

\author{Annick Suzor-Weiner}
\affiliation{Laboratoire de Photophysique Mol\'{e}culaire du CNRS, Universit\'{e} Paris-Sud, B\^{a}timent 210, F-91405 Orsay, France.}

\author{Eric Charron}
\affiliation{Laboratoire de Photophysique Mol\'{e}culaire du CNRS, Universit\'{e} Paris-Sud, B\^{a}timent 210, F-91405 Orsay, France.}

\date{\today}

\begin{abstract}
The ultrafast electronic and nuclear dynamics of H$_{2}$ laser-induced double ionization is studied using a time-dependent wave packet approach that goes beyond the fixed nuclei approximation. The double ionization pathways are analyzed by following the evolution of the total wave function during and after the pulse. The rescattering of the first ionized electron produces a coherent superposition of excited molecular states which presents a pronounced transient H$^{+}$H$^{-}$ character. This attosecond excitation is followed by field-induced double ionization and by the formation of short-lived autoionizing states which decay via double ionization. These two double ionization mechanisms may be identified by their signature imprinted in the kinetic-energy distribution of the ejected protons.
\end{abstract}
\maketitle

Laser-induced non-sequential double ionization of atoms has been extensively studied since its discovery in the early 1980s\,\cite{NSDI}. This strong field effect as well as high-order harmonic generation (HHG) in rare gases\,\cite{HOHG} gave birth to the so-called rescattering model which gives a unified picture of the atomic response\,\cite{RESCATTERING}. Moreover, this model constitutes the natural framework for the production of attosecond pulses and for the control of ionization in atoms and molecules\,\cite{ATTO}. On the molecular side, new effects in strong laser fields involve the coupling of the electronic and nuclear motions such as bond softening\,\cite{BS} or charge resonance enhanced ionization\,\cite{CREI}.  Recently, this coupling was shown to play a determinant role in HHG\,\cite{Lein05,Baker06,Wagner06}. In particular, Lein predicted that more intense harmonics are generated in heavier molecular isotopes\,\cite{Lein05}. This effect has been demonstrated experimentally using H$_2$, D$_2$, CH$_4$ and CD$_4$\,\cite{Baker06}. In addition, molecular vibrations were recently detected with a m{\AA} sensitivity using HHG in SF$_6$\,\cite{Wagner06}. Following a recent theoretical proposal\,\cite{Goll06}, ground state vibrational wave packets of D$_2$ where also mapped with 20\,m{\AA} sensitivity using selective depletion via tunnel ionization\,\cite{Ergler06}.

These recent advances show that nuclear motions are fast enough to exercise a decisive influence on the overall electronic response of the molecule in strong fields although the vibrational time scale only belongs to the femtosecond domain. Moreover, the hydrogen molecule exhibits the shortest vibrational period: 7.5\,fs. From the theoretical point of view it is therefore highly desirable to develop a unified and realistic theoretical framework where the nuclear dynamics is treated at the same quantum level as the electronic excitation. By treating the internuclear distance $R$ as a full quantum variable, we show that impulsive vibrational excitation also takes place in H$_{2}$. In addition, the rescattering dynamics is basically a two-electron problem which demands a two-active-electron approach. This approach allows for the description of the interplay between electronic rescattering and nuclear vibrations. Using a single-optical-cycle pulse, we show that the electron rescattering leads {\em during\/} the pulse to field-induced double ionization, and {\em after\/} the end of the pulse to auto-double ionization.

Using the splitting technique of the short-time propagator, we solve the time-dependent Schr\"odinger equation for H$_{2}$ in the linearly polarized classical electric field $\mathbi{E}(t)= E_{0}\,f(t)\cos(\omega t + \varphi)\;\hat{\mathbi{\!e}}$, where $\omega$ denotes the angular frequency, $E_{0}$ the field amplitude, and $\varphi$ the carrier-envelope offset phase. The pulse shape, of total duration $\tau$, is $f(t)=\sin^{2}(\pi t/\tau)$. The frequency $\omega$ corresponds to a central wavelength of $800$\,nm. The internuclear coordinate $\mathbi{R}$ is constrained along the field polarization vector $\hat{\mathbi{\!e}}$ and the two electrons, of coordinates $\mathbi{r}_i=z_i\;\hat{\mathbi{\!e}}$, are assumed to oscillate along the same axis. This one-dimensional approach was shown recently to represent faithfully most of the electronic dynamics\,\cite{Becker06}. The wave function $\Psi(R,z_1,z_2,t)$ of H$_{2}$ is taken initially as the ground state of the soft-Coulomb potential
\begin{eqnarray}
\label{eq:Vc}
V_{\mathrm{c}}(R,z_1,z_2) & = & 1/R + 
           \left[ (z_2 - z_1)^2 + \alpha^2(R) \right]^{-\frac{1}{2}} \nonumber
           \\
            & - &\sum_{\substack{i=1,2 \\ s=\pm 1}}
             \left[ (z_i + s\,R/2)^2+\beta^2(R) \right]^{-\frac{1}{2}}
\end{eqnarray}
which takes into account the electrostatic repulsion and attraction between all particles\,\cite{LF}. In order to mimic the dynamics of the real H$_{2}$ molecule, we have introduced two softening parameters, $\alpha(R)$ and $\beta(R)$, which vary slowly with the internuclear distance. These parameters have been adjusted at each internuclear distance to reproduce accurately the potential curves of H$_{2}$ and H$_{2}^{+}$ ground electronic states. All spectroscopic constants and ionization potentials of our model H$_{2}$ and H$_{2}^{+}$ molecules therefore match very closely the tabulated ones.

\begin{figure}[!t]
\centering
\includegraphics*[width=8.6cm,clip=true]{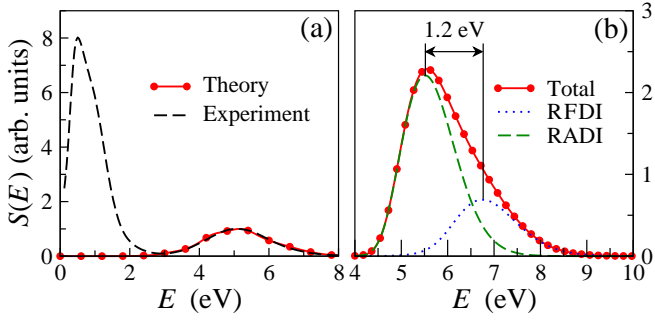}
\caption{(Color online) (a) Theoretical and experimental proton kinetic energy spectra at 4.5$\times$10$^{14}$\,W/cm$^{2}$. The measured pulse duration (FWHM) is 10$\pm 2\,$fs and the calculation is for 11.7\,fs. (b) Theoretical spectrum at 5$\times$10$^{14}$\,W/cm$^{2}$ and for $\tau=2.67$\,fs. The dotted and dashed lines correspond to recollision-induced field-assisted double ionization (RFDI) and to recollision-induced auto-double ionization (RADI).}
\label{fig:KER}
\end{figure}

During the pulse, the Hamiltonian includes the radiative coupling $(z_1+z_2)E(t)$ in the dipole length representation. The outgoing ionization flux is summed to extract the single and double ionization probabilities. For this purpose, we partition the plane $(z_1,z_2)$ in three regions $\Gamma_0$, $\Gamma_1$ and $\Gamma_2$ corresponding respectively to H$_{2}$, H$_{2}^{+}$, and H$_{2}^{2+}$. Double ionization occurs in the asymptotic region $\Gamma_2 \equiv \{|z_1|,|z_2|>z_{\mathrm{A}}\}$ with $z_{\mathrm{A}} = 20\,a_0 \gg R$. The single ionization region $\Gamma_1$ is defined as $\{|z_i|<z_{\mathrm{A}}$, $|z_j|>z_{\mathrm{A}}\}$, and the neutral H$_{2}$ molecule is found in the region $\Gamma_0\equiv\{|z_1|,|z_2| \leqslant z_{\mathrm{A}}\}$. The total wave function $\Psi(R,z_1,z_2,t)$ is propagated on a spatial grid of $(512)^3$ points with $z_{\mathrm{max}} = 100\,a_0$, $R_{\mathrm{max}} = 10\,a_0$ and $\delta t = 1\,$as. A simple mapping relates the proton kinetic energy distribution $S(E,t)$ to $P_2(R,t) = \iint_{\Gamma_2} |\Psi(R,z_1,z_2,t)|^2 \, dz_{1} \, dz_{2}$ using $E=1/(2R)$ and the requirement of particle conservation $P_2(R,t)\,dR=S(E,t)\,dE$. This energy distribution is accumulated over the entire time propagation to obtain the kinetic energy release spectrum $S(E)$ which is measured in experiments. Fig.\,\ref{fig:KER}(a) shows that a good agreement is obtained for the Coulomb peak in the energy range $3\,\mathrm{eV} \leqslant E \leqslant 8\,\mathrm{eV}$ between this model and the experiment\,\cite{SACLAY}. The low energy peak ($E \leqslant 3\,\mathrm{eV}$), which results from the photodissociation of H$_{2}^{+}$ in H\,+\,H$^{+}$, is not reproduced by this Coulomb mapping.

\begin{figure}[!t]
\centering
\includegraphics[width=8.6cm,clip=true]{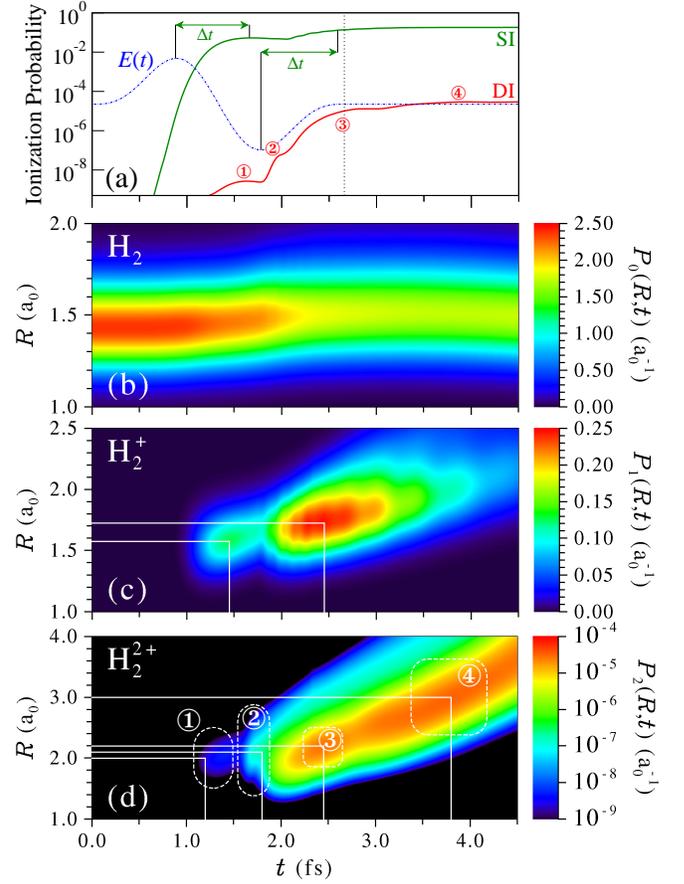}
\caption{(Color online) (a): Single (SI) and double (DI) ionization probabilities as a function of time at 5$\times$10$^{14}$\,W/cm$^{2}$ and for $\tau=2.67$\,fs. The electric field $E(t)$ is superimposed on this graph. The vertical line marks the end of the pulse. (b), (c) and (d): Nuclear probability distributions $P_k(R,t)= \iint_{\Gamma_k} |\Psi(R,z_1,z_2,t)|^2 \, dz_{1} \, dz_{2}$ as a function of internuclear distance $R$ and time $t$ in the three regions $\Gamma_k$ corresponding to H$_{2}$ (b), H$_{2}^{+}$ (c) and H$_{2}^{2+}$ (d).}
\label{fig:Pionization}
\end{figure}

A simplified analysis of the various double ionization mechanisms can be performed using the shortest pulse containing a complete oscillation of the field, with $\tau = 2 \pi / \omega = 2.67\,$fs and $\varphi=-\pi/2$. The associated electric field $E(t)$, which is shown Fig.\,\ref{fig:Pionization}(a) as a thin dotted curve, presents two symmetric maxima pointing in opposite directions. The time variations of the single (SI) and double (DI) ionization probabilities are also shown Fig.\,\ref{fig:Pionization}(a) for a peak intensity of 5$\times$10$^{14}$\,W/cm$^{2}$. Single and double ionization take place by successive bursts. The probability of single ionization is simply rising just after each maximum of the field, and the ionized electron reaches the asymptotic region in $\Delta t \sim 700\,$as. The evolution of the double ionization probability is more complex. Indeed, double ionization takes place in four successive bursts which differ considerably by their magnitudes and time scales. Since the first appearance of double ionization (label \ding{172}, Fig.\,\ref{fig:Pionization}(a)) follows the first maximum of the field by about 700\,as, it is clearly due to laser-induced direct double ionization. The second burst (\ding{173}) just follows the second maximum of the field and thus probably arises from single ionization of the H$_{2}^{+}$ molecular ions already produced. Finally, we will show in the following that the two last and dominant pathways to double ionization (\ding{174},\ding{175}) are induced by recollision and can be attributed to laser-assisted and to autoionization processes.

Fig.\,\ref{fig:Pionization}(b), which represents the evolution of the wave packet in the $\Gamma_0$ region, shows that, even for a pulse of total duration $\tau=2.67\,$fs, a significant nuclear motion takes place in H$_{2}$. At about 1.2\,fs, the neutral molecule stretches from the equilibrium internuclear distance $1.4\,a_0$ to $1.5\,a_0$. This vibrational wave packet, coherent superposition of a few vibrational levels, arises from a small and early {\em lochfra\ss\/} effect\,\cite{Goll06} which is amplified by a subsequent bond-softening like dynamics. The competition between these two mechanisms was monitored very recently in D$_2$\,\cite{Ergler06}.

The nuclear dynamics taking place during single and double ionization is even more significant. Figures\,\ref{fig:Pionization}(c) and \ref{fig:Pionization}(d) show that the chemical bond stretches up to $1.7\,a_0$ when the molecule is singly ionized and up to $3\,a_0$ during double ionization. The first three bursts of double ionization are identified in Fig.\,\ref{fig:Pionization}(d) by the labels \ding{172}, \ding{173} and \ding{174}. These three processes take place at similar internuclear distances, $2-2.2\,a_0$, close to the minimum of the potential associated to the ground electronic state of H$_{2}^{+}$ ($2\,a_0$). The last mechanism (label \ding{175}) is taking place much later, $t \sim 3-4.5\,$fs, after the end of the pulse, and around $3\,a_0$ This distance is far beyond the equilibrium distances of H$_{2}$ and H$_{2}^{+}$. In this reaction path, a dissociation process is therefore already taking place during the second stage of ionization, before Coulomb explosion.

\begin{figure}[!t]
\centering
\includegraphics[width=8.6cm,clip=true]{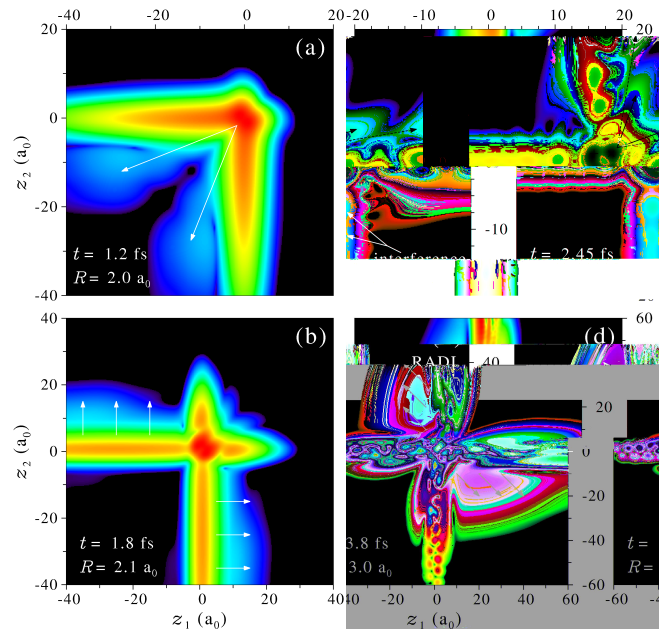}
\caption{(Color online) Electronic probability distribution $|\Psi(R,z_{1},z_{2},t)|^{2}$ (log scale) at different times and internuclear distances. The white arrows represent the photoelectron flux. The panels a-d correspond to the points 1-4 in Fig.\,\ref{fig:Pionization}(d).}
\label{fig:Pelectron}
\end{figure}

A clear identification of the four mechanisms giving rise to Coulomb explosion of H$_{2}$ is obtained by analyzing the electronic dynamics shown in Fig.\,\ref{fig:Pelectron} at specific times and internuclear distances. The times $t$ and distances $R$ we have chosen correspond to the four labels given in Fig.\,\ref{fig:Pionization}(d). In agreement with the previous discussion, we observe that laser-induced {\em direct} double ionization takes place around 1.2\,fs (white arrows in Fig.\,\ref{fig:Pelectron}(a)). Indeed, the two electrons escape {\em simultaneously} from the nuclear attraction in the same direction ($z_1$ and $z_2<0$), opposite to the direction of the electric field. In addition, the electronic repulsion quickly washes out the electronic density around $z_1 = z_2$. The curvature of the two pathways seen in this figure indicates that an energy exchange is taking place, inducing an acceleration of one of the electrons to the detriment of the other. Similar effects were seen earlier in the context of multi-electron ionization of atoms\,\cite{Eichmann}. The electronic probability distribution of Fig.\,\ref{fig:Pelectron}(b) also confirms that {\em sequential} double ionization occurs around $1.8\,$fs. In this case, the first electron has already left the molecular ion in the direction $z < 0$ when the second electron is ejected. The driving electric field has then reversed sign, and this second electron is leaving the nuclei in the opposite direction.

The nature of the two dominant pathways to double ionization is finally enlightened by the snapshots of the electronic density given in the last two panels of Fig.\,\ref{fig:Pelectron}. When the electric field changes sign ($t > 1.33\,$fs), the part of the electronic wave function which is located in the single ionization region $\Gamma_1$, and which therefore represents an ionized electron far from the bound electron of H$_{2}^{+}$, starts to be driven back towards the nuclei. This recollision process induces an interference between the incoming (recollision) and outgoing (ionization) electron waves. This interference is easily seen in the oscillatory behavior of the electronic density in Fig.\,\ref{fig:Pelectron}(c). The two peaks identified in this figure, which are separated by $\Delta z \sim 5\,a_0$, will recollide successively with the electron that remained near the nuclei in a time interval of about 115\,as. Because of this interference pattern, the recollision takes place as a very fast succession of periodic electron impacts. Each impact induces a very specific electronic excitation in the central region where $|z_1|$ and $|z_2| \leqslant R$. Indeed, in the vicinity of the two nuclei the electronic wave function does not show anymore the usual double peak structure of H$_{2}$'s ground electronic state. This structure is now replaced by a dynamical correlated motion of the two electrons around the four crosses shown in this figure. The two `$+$' signs correspond to the two attractive combinations where each electron is located on a different nucleus. The two `$\times$' signs correspond to the two ionic combinations H$^{+}$H$^{-}$ and H$^{-}$H$^{+}$. Around the internuclear distance $2.2\,a_0$, the energy gap between the ionic state and the ground electronic state of H$_{2}$ can be estimated from the energy difference between the $B\,^{1}\Sigma_{u}^{+}$ and the $X\,^{1}\Sigma_{g}^{+}$ potential curves of H$_{2}$ as $\Delta \sim 0.37\,E_{\textrm{h}}$. Considering the classical ponderomotive energy $U_p = e^{2}E_{\mathrm{max}}^{2}/4m\omega^2 \sim 0.46\,E_{\textrm{h}}$ for the present maximum value of the field $E_{\mathrm{max}} \sim 0.08\,E_{\textrm{h}}/e\/a_0$, the H$^{+}$H$^{-}$ spectral range is clearly accessible following the recollision process. In addition, recent studies\,\cite{Ionic} have shown that one of these ionic states is stabilized by strong fields, such that the ionic and the ground electronic states of H$_{2}$ may experience an avoided crossing in an adiabatic picture. As a consequence, the ionic configuration H$^{+}$H$^{-}$ becomes the instantaneous ground state of the system for large field amplitudes. We find here that this effect, which was first demonstrated in a two-electron model of H$_{2}$ with frozen nuclei\,\cite{Ionic}, survives in a more realistic approach which takes into account the dissociative dynamics of the molecule. For large field values, and because of the recollision, the two electrons are thus forced to localize on the same nucleus. Every 115\,as, this ionic state is transiently populated by recollision, and in this configuration the two electrons can easily escape by tunnel ionization. We therefore observe a very fast succession of Recollision-induced Field-assisted Double Ionization (RFDI, see the four double ionization arrows in Fig.\,\ref{fig:Pelectron}(c)).

As mentioned previously, the last efficient pathway to double ionization takes place after the end of the pulse, and therefore results from an autoionization process. This interpretation is confirmed by the snapshot of the electronic wave packet shown in Fig.\,\ref{fig:Pelectron}(d). For $t>3.0$\,fs, the two electrons escape in opposite directions since the ionization is now mainly driven by the electron repulsion. This ionization pathway, as well as the RFDI pathway of Fig.\,\ref{fig:Pelectron}(c), are suppressed when we introduce absorbing boundaries for the rescattering flux. The origin of both ionization channels is therefore to be found in the electron recollision. We therefore name the post-pulse ionization process ``Recollision-induced Auto-Double Ionization'', or RADI. Additional calculations performed with fixed nuclei show that RADI also takes place when the vibrational dynamics is frozen. This process is therefore not induced by a vibrational coupling\,\cite{VibAuto}, but is certainly due to transiently populated \textit{dressed} molecular states lying above the \textit{field-free} double ionization threshold. When the field is suddenly switched off, these highly excited states decay rapidly by double ionization. This interpretation is confirmed by the fact that RADI disappears for pulse durations longer than 4 optical cycles.

Finally Fig.\,\ref{fig:KER}(b) reports the contribution of these last two ionization channels to the proton kinetic energy spectrum. Since autoionization occurs at larger internuclear distances, the corresponding proton spectrum (dashed line) appears at lower energies than the RFDI proton peak (dotted line). The total spectrum, shown as a solid line with circles, includes the four double ionization pathways of Fig.\,\ref{fig:Pionization}(a). However, the direct and sequential contributions \ding{172} and \ding{173} remain very weak in comparison with the RFDI and RADI mechanisms. The asymmetry of the fragment kinetic energy release spectrum therefore reveals the presence of these last two pathways, induced by the rescattering process. Additional calculations show that the competition between RFDI and RADI is mainly controlled by the pulse duration and intensity. RFDI is favored with longer pulses and\,/\,or higher intensities.

To conclude, we have presented a theoretical model for the study of the electronic and nuclear dynamics of diatomic molecules in intense laser fields. This model can be adapted to a particular molecule by a simple adjustment of two softening functions $\alpha(R)$ and $\beta(R)$. We have applied it to the simplest molecular target H$_{2}$. We have shown that at moderate intensities molecular double ionization is mainly driven by the rescattering process. This process induces a specific transient ionic excitation of the molecule which leads to Recollision-induced Field-assisted Double Ionization (RFDI) and to Recollision-induced Auto-Double Ionization (RADI). These two double ionization mechanisms initiate the Coulomb explosion of the molecule at different internuclear distances, an effect which can be observed in the fragment kinetic energy release spectrum. These mechanisms could also be investigated in the time domain using attosecond EUV pulses. Indeed, a time-delayed attosecond pulse may induce an efficient single ionization of H$_{2}$ in the energy range $\hbar\omega \sim 20\,$eV, thus washing out the electron correlation in a few tens of attoseconds. Controlling the time-delay between the IR femtosecond pump and the EUV attosecond probe should therefore make it possible to reveal experimentally these two mechanisms by removing them one after the other.

\begin{acknowledgments}
We acknowledge HPC facilities of IDRIS-CNRS (06-1459) and financial support from ACI Photonique Physique Attoseconde and from CEA (LRC-DSM 05-33).
\end{acknowledgments}


\end{document}